\documentclass[fleqn,usenatbib]{mnras}

\usepackage{newtxtext,newtxmath}

\usepackage[T1]{fontenc}

\DeclareRobustCommand{\VAN}[3]{#2}
\let\VANthebibliography\thebibliography
\def\thebibliography{\DeclareRobustCommand{\VAN}[3]{##3}\VANthebibliography}

\usepackage{graphicx}	
\usepackage{amsmath}	

\title[BOAT GRB without Jet Break]{TeV Afterglow of BOAT GRB without Jet Break}

\author[Y. Kusafuka et al.]{
Yo Kusafuka,$^{1}$\thanks{E-mail: kusafuka@icrr.u-tokyo.ac.jp}
Katsuaki Asano$^{1}$
\\
$^{1}$Institute for Cosmic Ray Research, The University of Tokyo, 5-1-5 Kashiwanoha, Kashiwa, Chiba 277-8582, Japan
}

\date{Accepted XXX. Received YYY; in original form ZZZ}

\pubyear{\the\year{}}

\begin{document}
\label{firstpage}
\pagerange{\pageref{firstpage}--\pageref{lastpage}}
\maketitle

\begin{abstract}
We present a new model for the TeV afterglow of GRB 221009A.
The rapid increase of the TeV flux in the very early phase is reproduced by the magnetic acceleration. 
We consider the change in the radial structure of the circumstellar medium from homogeneous to wind-like to describe the breaks in the TeV light curve.
Our results imply a highly magnetized ejecta with a significantly thick width, making the deceleration time around 400 s for observers.
In our model, no early jet break is required.
\end{abstract}




\section{Introduction} \label{sec:intro}


The physical mechanism generating afterglow radiation in gamma-ray bursts (GRBs) is thought to be external shocks caused by the interaction between relativistic ejecta and the circumstellar medium (CSM). 
In addition to the usual synchrotron component, some GRB afterglows show very high-energy (VHE) gamma rays, which are thought to be produced by the synchrotron self-Compton (SSC) mechanism \citep{2001ApJ...548..787S,2001ApJ...552L..35Z,2008MNRAS.384.1483F}.
Thanks to imaging atmospheric Cherenkov telescopes (IACTs), the number of VHE afterglows has been increasing: GRB 180720B \citep{2019Natur.575..464A}, GRB 190114C \citep{2019Natur.575..455M}, GRB 190829A \citep{2021Sci...372.1081H}, and GRB 201216C \citep{2024MNRAS.527.5856A}. 

Recently, numerous TeV photons from the brightest-of-all-time (BOAT) GRB 221009A located at $z=0.151$ \citep{2023arXiv230207891M} 
were detected by the Large High Altitude Air Shower Observatory (LHAASO) \citep{2023Sci...380.1390L}. 
GRB 221009A was first detected by the Gamma-ray Burst Monitor (GBM) on aboard Fermi satellite ($T_0$: 2022 October 9 at 13:16:59.99 UTC) \citep{2023ApJ...952L..42L}. 
The prompt emission of the burst lasts more than 600 s. 
The standard duration of the temporal interval that contains 90\% of the total time-integrated photon flux in the 50–300 keV energy range is $T_{90}\sim300$ s \citep{2023ApJ...952L..42L}. 
Spectral analysis of prompt emission suggests dominance of the Poynting flux at the dissipation radius \citep{2023ApJ...947L..11Y,2024JHEAp..41...42Z}, which may imply a magnetically driven mechanism for launching the jet. 

The TeV light curve of GRB 221009A presents a sharp increase $\propto T^{14.9}$ at $\sim T_0 + 230$ s, turns into a relatively slow rise $\propto T^{1.8}$ until the peak at $\sim T_0 + 245$ s and then decays as $\propto T^{-1.1}$ with a steep decay at $\sim T_0 + 670$ s \citep{2023Sci...380.1390L}. 
The sharp increase might be explained by the effects of $\gamma \gamma$ annihilation due to interaction with the internal photons or the prompt MeV photons \citep{2023arXiv231113671Z,2024ApJ...961L...6G}. 
In the previous models, the peak at $\sim T_0 + 245$ s is considered as a deceleration time scale, and the break in the light curve around $\sim T_0 + 670$ s is believed to be signature of the jet break \citep{2023Sci...380.1390L,2023MNRAS.522L..56S,2024ApJ...962..115R}. In such models, the jet opening angle is required to be exceptionally narrow as $\sim0.01$ rad. 

The late phase afterglow of GRB 221009A data from radio to GeV bands were also obtained through multiwavelength observations \citep{2023NatAs...7..986B,2023ApJ...948L..12K,2023ApJ...946L..23L,2023Sci...380.1390L}. 
Using these data, the CSM structure has been discussed.
\citet{2023ApJ...948L..12K,2023SciA....9I1405O,2023MNRAS.522L..56S} claim homogeneous CSM; while \citet{2023ApJ...946L..23L,2023ApJ...947...53R,2023MNRAS.524L..78G} suggest a wind-like medium. 
\citet{2024ApJ...962..115R} point out that the CSM structure may be inverted, and then they can reproduce well from the early to the late phase afterglow with a structured jet.

Recently, \citet{2024MNRAS.536.1822K} provided a detailed study of the effects of magnetization and thickness of the ejecta.
Using this model, here we try to propose an alternative explanation for GRB 221009A.

\section{Model and Method} \label{sec:model}

To reproduce the multi-wavelength afterglow light curves in GRB 221009A, especially for the early stage, we adopt the formulae of the forward shock evolution obtained by \citet{2024MNRAS.536.1822K}. The formulas take into account the magnetization and the finite thickness of the ejecta \citep[see,][]{2000ApJ...542..819K,2009A&A...494..879M,2014MNRAS.442.3495V,2023MNRAS.526..512K}. 
Initially, the ejecta is accelerated by magnetic pressure \citep[impulsive acceleration;][]{2010PhRvE..82e6305L,2011MNRAS.411.1323G}. Then, at the ignition time of the reverse shock, the ejecta and the forward shock transit to the coasting phase. As the density of the ejecta decreases as $n_{\rm ej}\propto r^{-2}$, the Lorentz factor of the forward shock $\Gamma_{\rm FS}$ gradually starts to decrease. Even after the reverse shock breakout, because of the finite thickness of the ejecta, the energy transfer from the ejecta to the shocked CSM continues, leading to a gradual deceleration. This transition phase lasts until the catch-up of the rarefaction wave, i.e. the edge of the ejecta, with the forward shock front. Then, the standard deceleration phase expressed by the Blandford-Mckee (BM) self-similar solution \citep{1976PhFl...19.1130B} is followed. 

The advantages of introducing the formulae in \citet{2024MNRAS.536.1822K} are as follows. In the acceleration phase, the TeV flux can increase roughly by $\propto T^{15}$, which is consistent with the observed value $T^{14.9}$ for $T<T_0 + 230$ s \citep{2023Sci...380.1390L}.
Due to the finite thickness of the ejecta, the switching time $T_{\rm BM}$ to the BM solution for observers can be the observed break time at $T_0 + 670$ s. In this case, we do not need an extremely narrow jet as required in the jet break models.


We assume a top-hat jet with a finite thickness ejected from the central engine. The duration of the prompt emission of GRB 221009A is $\sim 600$ s \citep{2023ApJ...946L..31B}, while the main pulse lasted $\sim 100$ s \citep{
2024arXiv240904580A}. The thickness of the ejecta $\Delta_0$ would be $\gtrsim c \times 100$ s. 


To reproduce the complicated behavior of the TeV gamma-ray afterglow, following \citet{2024ApJ...962..115R}, we consider an ``inverted'' CSM structure: a homogeneous medium near the progenitor star and change to a wind-like medium at a specific radius $r_{\rm br}$,
\begin{equation}
    n_{\rm CSM}(r)=3\times10^{35}{\rm{cm}}^{-1} A
    \left\{
    \begin{array}{ll}
    r_{\rm br}^{-2}\ \ \ \ r\leqq r_{\rm br} \\
    r^{-2}\ \ \ \ r>r_{\rm br}
    \end{array}
    \right.,
    \label{eq:den}
\end{equation}
where $A=(\dot{M}/10^{-5}\ M_{\sun} {\rm yr}^{-1})(v_{\rm wind}/10^3\ {\rm km}{\rm s}^{-1})^{-1}$ represents a mass-loss rate of the progenitor star by the 
stellar wind \citep{2018MNRAS.478..110S,2024ApJ...960...70M}.

\begin{table*}
 \caption{Model fitting parameters. From the left, the maximum Lorentz factor of the forward shock $\Gamma_{\rm max}$, CSM break radius $r_{\rm br}$, isotropic-equivalent energy $E_{\rm iso}$, jet opening angle $\theta_0$, typical ejecta magnetization at the end of the acceleration phase $\sigma_{\rm RS}$, initial radial width of the ejecta $\Delta_0$, wind density parameter $A$, spectral index of accelerated electrons at injection $p$, accelerated electron energy fraction $\epsilon_{\rm e}$, magnetic energy fraction $\epsilon_B$, and number fraction of accelerated electrons $f_{\rm e}$.}
 \label{table:models}
 \centering
  \begin{tabular}{ccccccccccc}
   \hline
   $\Gamma_{\rm max}$ & $r_{\rm br}$ [cm] & $E_{\rm iso}$ [erg] & $\theta_0
   $ [rad] & $\sigma_{\rm RS}$ & $\Delta_{0}/c$ [s] & $A$ [${\rm cm}^{-1}$] & $p$ & $\epsilon_e$ & $\epsilon_B$ & $f_{\rm e}$ \\
   \hline \hline
   $530$ & $1.2\times10^{17}$ & $4.0\times10^{55}$ & $3.0\times10^{-2}$ & $23$ & $200$ & $4.8\times10^{-2}$ & $2.45$ & $3.0\times10^{-2}$ & $4.6\times10^{-5}$& $1.0$ \\
   \hline
  \end{tabular}
\end{table*}


The model parameters and their definition \citep[see][for details]{2024MNRAS.536.1822K} are summarized in Table \ref{table:models}.
Initially, the bulk Lorentz factor evolves as $\Gamma \propto r^{1/3} \propto T$.
For highly magnetized ejecta, the reverse shock crossing time is significantly short \citep{2005ApJ...628..315Z,2009A&A...494..879M,2023MNRAS.526..512K,2024MNRAS.536.1822K},
so that we only consider the forward shock dynamics in the stratified medium.
Since the CSM number density is relatively high compared to the ejecta number density, the transition phase starts just after the end of the acceleration phase without the coasting phase. 
The forward shock Lorentz factor in the transition phase evolves as \citep{2024MNRAS.536.1822K} 
\begin{equation}
    \Gamma_{\rm FS}=\frac{\Gamma_{\rm max}}{\left[ 1+2\Gamma_{\rm max}\sqrt{\frac{n_{\rm CSM}}{n_{\rm ej}(1+\sigma_{\rm RS})}} \right]^{1/2}},
    \label{eq:Gamma_trans}
\end{equation}
where $n_{\rm ej}$ is the number density of the unshocked ejecta, $\sigma_{\rm RS}$ is the magnetization of the shocked ejecta, and $\Gamma_{\rm max}$ is the maximum Lorentz factor just after the acceleration phase. 
In our case, the second term in the denominator becomes dominant, so that we obtain
\begin{equation}
    \Gamma_{\rm FS}\propto\left(r^2n_{\rm CSM}\right)^{-1/4}\propto
    \left\{
    \begin{array}{ll}
    r^{-1/2} \propto T^{-1/4}\ \ \ \ r\leqq r_{\rm br} \\
    r^0 \propto T^{0}\ \ \ \ \ \ \ \ r>r_{\rm br}
    \end{array}
    \right..
    \label{eq:Gamma_CSM}
\end{equation}
The continuous energy injection from the ejecta in the transition phase in the wind medium leads to maintaining the Lorentz factor like the coasting phase.

\begin{figure}
\includegraphics[width=\columnwidth]{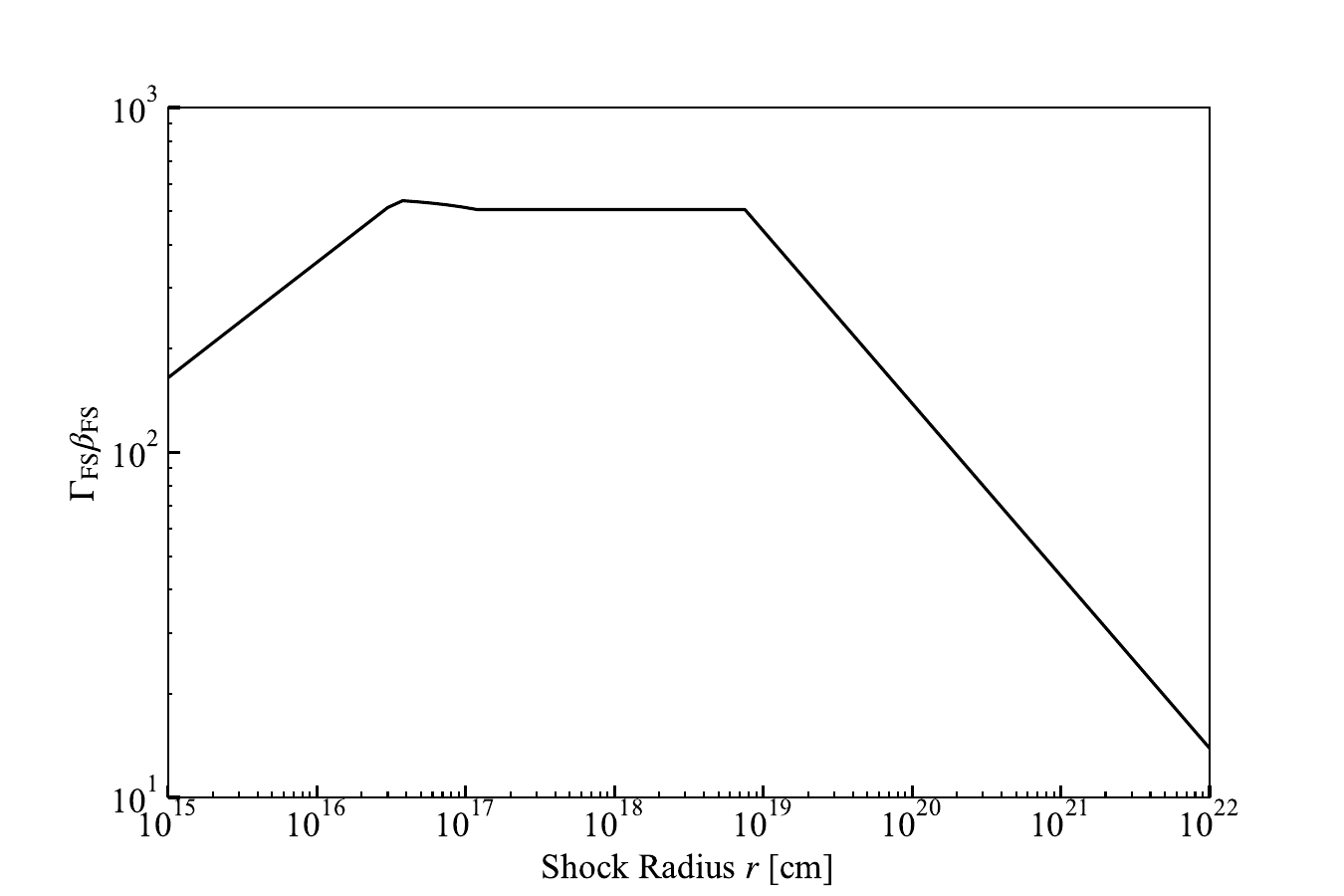}
\caption{The time evolution of the Lorentz factor just behind of the forward shock. The dynamics has 4 phases: 1. Acceleration phase in the homogeneous CSM, 2. Transition phase in homogeneous CSM, 3. Transition phase in the wind-like CSM, 4. Deceleration phase in the wind-like CSM. 
\label{fig:FS}}
\end{figure}

Given the width $\Delta_0$, we can determine when the rarefaction wave catches up to the forward shock front. After that, the Blandford-McKee self-similar solution in the wind-like medium $\Gamma_{\rm FS}\propto r^{-1/2} \propto T^{-1/4}$ \citep{1976PhFl...19.1130B} is adopted. The entire evolution of $\Gamma_{\rm FS}$ is shown in Figure \ref{fig:FS}.
The maximum Lorentz factor for the forward shock is $530$, which is almost comparable to the values reported in previous research \citep{2023Sci...380.1390L,2023SciA....9I1405O,2023MNRAS.522L..56S,2023ApJ...947...53R,2024ApJ...962..115R}.

Using the $\Gamma_{\rm FS}$ evolution in  Figure \ref{fig:FS} and the density profile given by Eq. (\ref{eq:den}), we calculate the evolution of synchrotron and Synchrotron Self-Compton (SSC) emission with synchrotron self-absorption and $\gamma \gamma$ absorption following the same calculation method described in \citet{2024MNRAS.536.1822K}.
Taking into account the relativistic Doppler and beaming effects, and the curvature effect of the emission surface, we produce the multi-wavelength lightcurves for observers.


\section{Results and Discussion} \label{sec:results}



\begin{figure*}
\includegraphics[width=15cm]{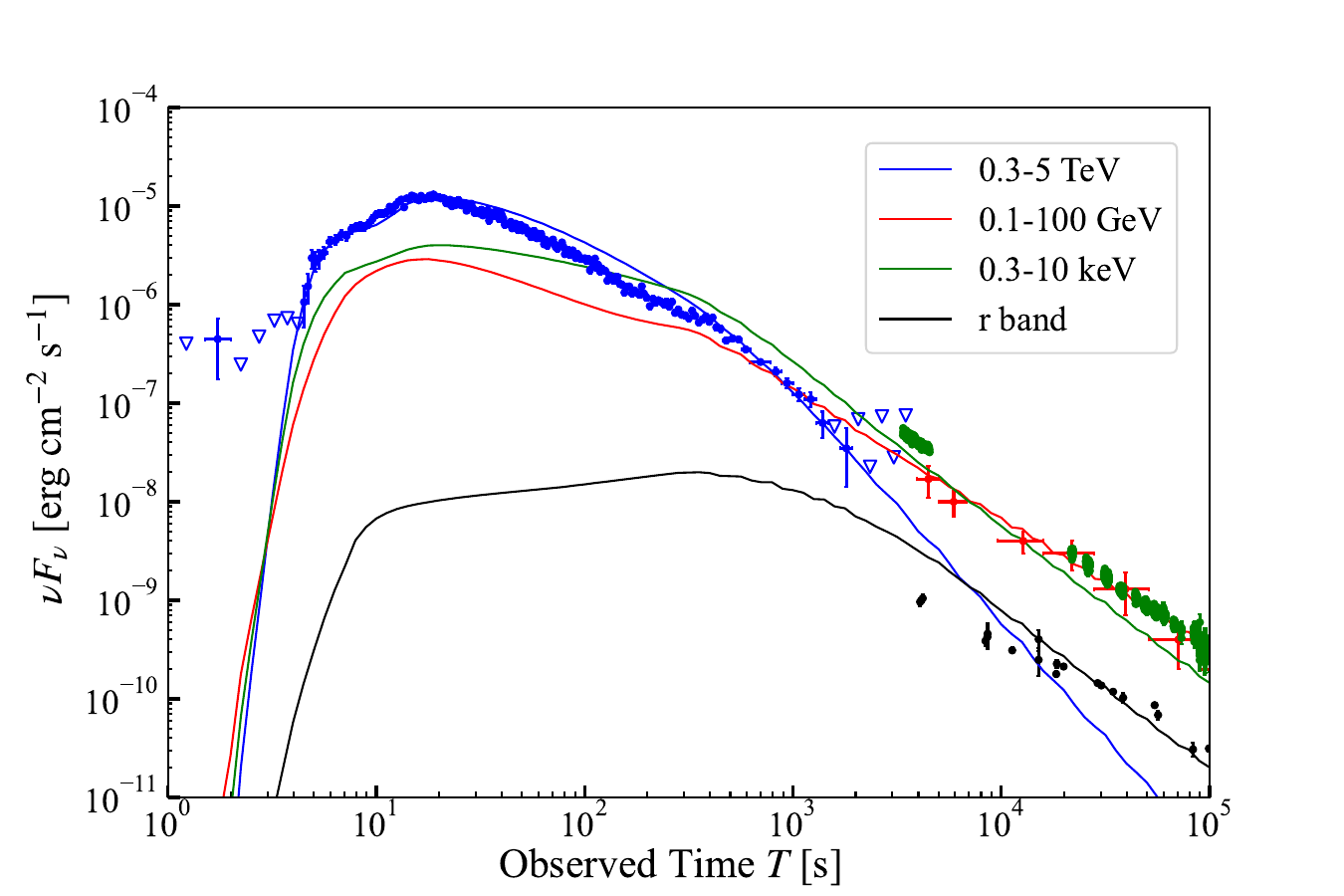}
\caption{The multi-wavelength data and our fitting light curve. 
Filled circles are data points and open triangles express the upper limit. The solid lines are our model results. 
\label{fig:LC}}
\end{figure*}




The results are shown in Figure \ref{fig:LC}, 
where the observed data are from \citet{2023Sci...380.1390L} for TeV gamma-ray, 
\citet{2024arXiv240904580A} for GeV gamma-ray, publicly available X-ray data from Swift-XRT, and \citet{2023ApJ...946L..23L} and \citet{2023ApJ...948L..12K} for optical emission. 
We applied the correction of the galactic extinction as $A_{\rm V}=4.1$ \citep{2011ApJ...737..103S} and the host correction as $A_{\rm V}=0.24$ \citep{2023ApJ...946L..23L}, but the late-time supernova and host galaxy contribution are not considered.
Note that the origin of the time is shifted from $T_0$ defined by {\it Fermi}-GBM.

Our model reproduces the complex TeV light curve well. The steep increase until $\sim 5$ s is due to the magnetic acceleration in the homogeneous medium. The gradual increase until $\sim 15$ s corresponds to the fast cooling regime 
in the transition phase. Our choice of the parameters makes the timing of the transition to the slow cooling and plunge into the wind-like region ($\sim 10$ s) occur almost simultaneously. 
The peak time ($\sim 15$ s) corresponds to the switching from the fast to slow cooling in the transition phase and the wind-like medium. The transition phase with the slow cooling leads to the gradual decay until $400$ s. The steep decline for $T>400$ s reflects the deceleration following the BM solution.

Our model also explains well the optical to GeV light curve. 
Although we have focused on the early afterglow here, the late-time radio afterglow in our model is brighter than the observed data. To reconcile with the late afterglow data, a time evolution of the microscopic parameters \citep[e.g.][]{2024ApJ...973L..44F} or additional components \citep[e,g.][]{2022MNRAS.513.1895R,2023MNRAS.522L..56S,2023ApJ...946L..23L,Sun_2024} may be required. However, such complicated models for the late phase are beyond the scope of this paper.





The opening angle $\theta_0$ should be large enough not to affect the early afterglow in our model. The jet break appears 
around $10^{7}$ s for our choice of $\theta_0=0.03$ rad ($=1.7^\circ$), which is consistent with the mean value $=2.5\pm1^\circ$ estimated by \citet{2018ApJ...859..160W}.
In this case, the beaming-corrected jet energy is estimated as $\sim2\times10^{52}$ erg. According to $E_{\rm iso}=4 \times 10^{55}$ erg, the efficiency of prompt emission becomes $\eta_\gamma\sim0.2$,
which is consistent with the typical prompt efficiency \citep[e.g.][]{2016MNRAS.461...51B}.

We expect the Cherenkov Telescope Array Observatory (CTAO) to verify if a very rapidly increasing feature in the TeV lightcurve is frequent or universal, which provides a hint for the magnetization and early evolution of the relativistic jets.

\section*{Acknowledgements}

The authors thankfully acknowledge the computer resources provided by the Institute for Cosmic Ray Research (ICRR), the University of Tokyo. 
This work is supported by the joint research program of ICRR, and JSPS KAKENHI Grant Numbers JP23KJ0692 (Y.K.), and JP22K03684, JP23H04899, 24H00025 (K.A.).

\section*{Data Availability}

The data underlying this article will be shared on reasonable request to the corresponding author.



\bibliographystyle{mnras}
\bibliography{example} 

\begin{thebibliography}{}
\makeatletter
\relax
\def\mn@urlcharsother{\let\do\@makeother \do\$\do\&\do\#\do\^\do\_\do\%\do\~}
\def\mn@doi{\begingroup\mn@urlcharsother \@ifnextchar [ {\mn@doi@} {\mn@doi@[]}}
\def\mn@doi@[#1]#2{\def\@tempa{#1}\ifx\@tempa\@empty \href {http://dx.doi.org/#2} {doi:#2}\else \href {http://dx.doi.org/#2} {#1}\fi \endgroup}
\def\mn@eprint#1#2{\mn@eprint@#1:#2::\@nil}
\def\mn@eprint@arXiv#1{\href {http://arxiv.org/abs/#1} {{\tt arXiv:#1}}}
\def\mn@eprint@dblp#1{\href {http://dblp.uni-trier.de/rec/bibtex/#1.xml} {dblp:#1}}
\def\mn@eprint@#1:#2:#3:#4\@nil{\def\@tempa {#1}\def\@tempb {#2}\def\@tempc {#3}\ifx \@tempc \@empty \let \@tempc \@tempb \let \@tempb \@tempa \fi \ifx \@tempb \@empty \def\@tempb {arXiv}\fi \@ifundefined {mn@eprint@\@tempb}{\@tempb:\@tempc}{\expandafter \expandafter \csname mn@eprint@\@tempb\endcsname \expandafter{\@tempc}}}

\bibitem[\protect\citeauthoryear{{Abdalla} et~al.,}{{Abdalla} et~al.}{2019}]{2019Natur.575..464A}
{Abdalla} H.,  et~al., 2019, \mn@doi [\nat] {10.1038/s41586-019-1743-9}, \href {https://ui.adsabs.harvard.edu/abs/2019Natur.575..464A} {575, 464}

\bibitem[\protect\citeauthoryear{{Abe} et~al.,}{{Abe} et~al.}{2024}]{2024MNRAS.527.5856A}
{Abe} H.,  et~al., 2024, \mn@doi [\mnras] {10.1093/mnras/stad2958}, \href {https://ui.adsabs.harvard.edu/abs/2024MNRAS.527.5856A} {527, 5856}

\bibitem[\protect\citeauthoryear{{Axelsson} et~al.,}{{Axelsson} et~al.}{2024}]{2024arXiv240904580A}
{Axelsson} M.,  et~al., 2024, \mn@doi [arXiv e-prints] {10.48550/arXiv.2409.04580}, \href {https://ui.adsabs.harvard.edu/abs/2024arXiv240904580A} {p. arXiv:2409.04580}

\bibitem[\protect\citeauthoryear{{Beniamini}, {Nava}  \& {Piran}}{{Beniamini} et~al.}{2016}]{2016MNRAS.461...51B}
{Beniamini} P.,  {Nava} L.,   {Piran} T.,  2016, \mn@doi [\mnras] {10.1093/mnras/stw1331}, \href {https://ui.adsabs.harvard.edu/abs/2016MNRAS.461...51B} {461, 51}

\bibitem[\protect\citeauthoryear{{Blandford} \& {McKee}}{{Blandford} \& {McKee}}{1976}]{1976PhFl...19.1130B}
{Blandford} R.~D.,  {McKee} C.~F.,  1976, \mn@doi [Physics of Fluids] {10.1063/1.861619}, \href {https://ui.adsabs.harvard.edu/abs/1976PhFl...19.1130B} {19, 1130}

\bibitem[\protect\citeauthoryear{{Bright} et~al.,}{{Bright} et~al.}{2023}]{2023NatAs...7..986B}
{Bright} J.~S.,  et~al., 2023, \mn@doi [Nature Astronomy] {10.1038/s41550-023-01997-9}, \href {https://ui.adsabs.harvard.edu/abs/2023NatAs...7..986B} {7, 986}

\bibitem[\protect\citeauthoryear{{Burns} et~al.,}{{Burns} et~al.}{2023}]{2023ApJ...946L..31B}
{Burns} E.,  et~al., 2023, \mn@doi [\apjl] {10.3847/2041-8213/acc39c}, \href {https://ui.adsabs.harvard.edu/abs/2023ApJ...946L..31B} {946, L31}

\bibitem[\protect\citeauthoryear{{Fan}, {Piran}, {Narayan}  \& {Wei}}{{Fan} et~al.}{2008}]{2008MNRAS.384.1483F}
{Fan} Y.-Z.,  {Piran} T.,  {Narayan} R.,   {Wei} D.-M.,  2008, \mn@doi [\mnras] {10.1111/j.1365-2966.2007.12765.x}, \href {https://ui.adsabs.harvard.edu/abs/2008MNRAS.384.1483F} {384, 1483}

\bibitem[\protect\citeauthoryear{{Foffano}, {Tavani}  \& {Piano}}{{Foffano} et~al.}{2024}]{2024ApJ...973L..44F}
{Foffano} L.,  {Tavani} M.,   {Piano} G.,  2024, \mn@doi [\apjl] {10.3847/2041-8213/ad76a3}, \href {https://ui.adsabs.harvard.edu/abs/2024ApJ...973L..44F} {973, L44}

\bibitem[\protect\citeauthoryear{{Gao} \& {Zou}}{{Gao} \& {Zou}}{2024}]{2024ApJ...961L...6G}
{Gao} D.-Y.,  {Zou} Y.-C.,  2024, \mn@doi [\apjl] {10.3847/2041-8213/ad167d}, \href {https://ui.adsabs.harvard.edu/abs/2024ApJ...961L...6G} {961, L6}

\bibitem[\protect\citeauthoryear{{Gill} \& {Granot}}{{Gill} \& {Granot}}{2023}]{2023MNRAS.524L..78G}
{Gill} R.,  {Granot} J.,  2023, \mn@doi [\mnras] {10.1093/mnrasl/slad075}, \href {https://ui.adsabs.harvard.edu/abs/2023MNRAS.524L..78G} {524, L78}

\bibitem[\protect\citeauthoryear{{Granot}, {Komissarov}  \& {Spitkovsky}}{{Granot} et~al.}{2011}]{2011MNRAS.411.1323G}
{Granot} J.,  {Komissarov} S.~S.,   {Spitkovsky} A.,  2011, \mn@doi [\mnras] {10.1111/j.1365-2966.2010.17770.x}, \href {https://ui.adsabs.harvard.edu/abs/2011MNRAS.411.1323G} {411, 1323}

\bibitem[\protect\citeauthoryear{{H.~E.~S.~S. Collaboration} et~al.,}{{H.~E.~S.~S. Collaboration} et~al.}{2021}]{2021Sci...372.1081H}
{H.~E.~S.~S. Collaboration} et~al., 2021, \mn@doi [Science] {10.1126/science.abe8560}, \href {https://ui.adsabs.harvard.edu/abs/2021Sci...372.1081H} {372, 1081}

\bibitem[\protect\citeauthoryear{{Kann} et~al.,}{{Kann} et~al.}{2023}]{2023ApJ...948L..12K}
{Kann} D.~A.,  et~al., 2023, \mn@doi [\apjl] {10.3847/2041-8213/acc8d0}, \href {https://ui.adsabs.harvard.edu/abs/2023ApJ...948L..12K} {948, L12}

\bibitem[\protect\citeauthoryear{{Kobayashi} \& {Sari}}{{Kobayashi} \& {Sari}}{2000}]{2000ApJ...542..819K}
{Kobayashi} S.,  {Sari} R.,  2000, \mn@doi [\apj] {10.1086/317021}, \href {https://ui.adsabs.harvard.edu/abs/2000ApJ...542..819K} {542, 819}

\bibitem[\protect\citeauthoryear{Kusafuka \& Asano}{Kusafuka \& Asano}{2024}]{2024MNRAS.536.1822K}
Kusafuka Y.,  Asano K.,  2024, \mn@doi [MNRAS] {10.1093/mnras/stae2734}, 536, 1822

\bibitem[\protect\citeauthoryear{{Kusafuka}, {Asano}, {Ohmura}  \& {Kawashima}}{{Kusafuka} et~al.}{2023}]{2023MNRAS.526..512K}
{Kusafuka} Y.,  {Asano} K.,  {Ohmura} T.,   {Kawashima} T.,  2023, \mn@doi [\mnras] {10.1093/mnras/stad2804}, \href {https://ui.adsabs.harvard.edu/abs/2023MNRAS.526..512K} {526, 512}

\bibitem[\protect\citeauthoryear{{LHAASO Collaboration} et~al.,}{{LHAASO Collaboration} et~al.}{2023}]{2023Sci...380.1390L}
{LHAASO Collaboration} et~al., 2023, \mn@doi [Science] {10.1126/science.adg9328}, \href {https://ui.adsabs.harvard.edu/abs/2023Sci...380.1390L} {380, 1390}

\bibitem[\protect\citeauthoryear{{Laskar} et~al.,}{{Laskar} et~al.}{2023}]{2023ApJ...946L..23L}
{Laskar} T.,  et~al., 2023, \mn@doi [\apjl] {10.3847/2041-8213/acbfad}, \href {https://ui.adsabs.harvard.edu/abs/2023ApJ...946L..23L} {946, L23}

\bibitem[\protect\citeauthoryear{{Lesage} et~al.,}{{Lesage} et~al.}{2023}]{2023ApJ...952L..42L}
{Lesage} S.,  et~al., 2023, \mn@doi [\apjl] {10.3847/2041-8213/ace5b4}, \href {https://ui.adsabs.harvard.edu/abs/2023ApJ...952L..42L} {952, L42}

\bibitem[\protect\citeauthoryear{{Lyutikov}}{{Lyutikov}}{2010}]{2010PhRvE..82e6305L}
{Lyutikov} M.,  2010, \mn@doi [\pre] {10.1103/PhysRevE.82.056305}, \href {https://ui.adsabs.harvard.edu/abs/2010PhRvE..82e6305L} {82, 056305}

\bibitem[\protect\citeauthoryear{{MAGIC Collaboration} et~al.,}{{MAGIC Collaboration} et~al.}{2019}]{2019Natur.575..455M}
{MAGIC Collaboration} et~al., 2019, \mn@doi [\nat] {10.1038/s41586-019-1750-x}, \href {https://ui.adsabs.harvard.edu/abs/2019Natur.575..455M} {575, 455}

\bibitem[\protect\citeauthoryear{{Malesani} et~al.,}{{Malesani} et~al.}{2023}]{2023arXiv230207891M}
{Malesani} D.~B.,  et~al., 2023, \mn@doi [arXiv e-prints] {10.48550/arXiv.2302.07891}, \href {https://ui.adsabs.harvard.edu/abs/2023arXiv230207891M} {p. arXiv:2302.07891}

\bibitem[\protect\citeauthoryear{{Matsuoka}, {Kimura}, {Maeda}  \& {Tanaka}}{{Matsuoka} et~al.}{2024}]{2024ApJ...960...70M}
{Matsuoka} T.,  {Kimura} S.~S.,  {Maeda} K.,   {Tanaka} M.,  2024, \mn@doi [\apj] {10.3847/1538-4357/ad096c}, \href {https://ui.adsabs.harvard.edu/abs/2024ApJ...960...70M} {960, 70}

\bibitem[\protect\citeauthoryear{{Mimica}, {Giannios}  \& {Aloy}}{{Mimica} et~al.}{2009}]{2009A&A...494..879M}
{Mimica} P.,  {Giannios} D.,   {Aloy} M.~A.,  2009, \mn@doi [\aap] {10.1051/0004-6361:200810756}, \href {https://ui.adsabs.harvard.edu/abs/2009A&A...494..879M} {494, 879}

\bibitem[\protect\citeauthoryear{{O'Connor} et~al.,}{{O'Connor} et~al.}{2023}]{2023SciA....9I1405O}
{O'Connor} B.,  et~al., 2023, \mn@doi [Science Advances] {10.1126/sciadv.adi1405}, \href {https://ui.adsabs.harvard.edu/abs/2023SciA....9I1405O} {9, eadi1405}

\bibitem[\protect\citeauthoryear{{Ren}, {Wang}, {Zhang}  \& {Dai}}{{Ren} et~al.}{2023}]{2023ApJ...947...53R}
{Ren} J.,  {Wang} Y.,  {Zhang} L.-L.,   {Dai} Z.-G.,  2023, \mn@doi [\apj] {10.3847/1538-4357/acc57d}, \href {https://ui.adsabs.harvard.edu/abs/2023ApJ...947...53R} {947, 53}

\bibitem[\protect\citeauthoryear{{Ren}, {Wang}  \& {Dai}}{{Ren} et~al.}{2024}]{2024ApJ...962..115R}
{Ren} J.,  {Wang} Y.,   {Dai} Z.-G.,  2024, \mn@doi [\apj] {10.3847/1538-4357/ad1bcd}, \href {https://ui.adsabs.harvard.edu/abs/2024ApJ...962..115R} {962, 115}

\bibitem[\protect\citeauthoryear{{Rhodes}, {van der Horst}, {Fender}, {Aguilera-Dena}, {Bright}, {Vergani}  \& {Williams}}{{Rhodes} et~al.}{2022}]{2022MNRAS.513.1895R}
{Rhodes} L.,  {van der Horst} A.~J.,  {Fender} R.,  {Aguilera-Dena} D.~R.,  {Bright} J.~S.,  {Vergani} S.,   {Williams} D.~R.~A.,  2022, \mn@doi [\mnras] {10.1093/mnras/stac1057}, \href {https://ui.adsabs.harvard.edu/abs/2022MNRAS.513.1895R} {513, 1895}

\bibitem[\protect\citeauthoryear{{Sari} \& {Esin}}{{Sari} \& {Esin}}{2001}]{2001ApJ...548..787S}
{Sari} R.,  {Esin} A.~A.,  2001, \mn@doi [\apj] {10.1086/319003}, \href {https://ui.adsabs.harvard.edu/abs/2001ApJ...548..787S} {548, 787}

\bibitem[\protect\citeauthoryear{{Sato}, {Murase}, {Ohira}  \& {Yamazaki}}{{Sato} et~al.}{2023}]{2023MNRAS.522L..56S}
{Sato} Y.,  {Murase} K.,  {Ohira} Y.,   {Yamazaki} R.,  2023, \mn@doi [\mnras] {10.1093/mnrasl/slad038}, \href {https://ui.adsabs.harvard.edu/abs/2023MNRAS.522L..56S} {522, L56}

\bibitem[\protect\citeauthoryear{{Schlafly} \& {Finkbeiner}}{{Schlafly} \& {Finkbeiner}}{2011}]{2011ApJ...737..103S}
{Schlafly} E.~F.,  {Finkbeiner} D.~P.,  2011, \mn@doi [\apj] {10.1088/0004-637X/737/2/103}, \href {https://ui.adsabs.harvard.edu/abs/2011ApJ...737..103S} {737, 103}

\bibitem[\protect\citeauthoryear{{Sun} et~al.,}{{Sun} et~al.}{2024}]{Sun_2024}
{Sun} T.-R.,  et~al., 2024, \mn@doi [arXiv e-prints] {10.48550/arXiv.2409.17983}, \href {https://ui.adsabs.harvard.edu/abs/2024arXiv240917983S} {p. arXiv:2409.17983}

\bibitem[\protect\citeauthoryear{{Suzuki} \& {Maeda}}{{Suzuki} \& {Maeda}}{2018}]{2018MNRAS.478..110S}
{Suzuki} A.,  {Maeda} K.,  2018, \mn@doi [\mnras] {10.1093/mnras/sty999}, \href {https://ui.adsabs.harvard.edu/abs/2018MNRAS.478..110S} {478, 110}

\bibitem[\protect\citeauthoryear{{Wang}, {Zhang}, {Liang}, {Lu}, {Lin}, {Li}  \& {Li}}{{Wang} et~al.}{2018}]{2018ApJ...859..160W}
{Wang} X.-G.,  {Zhang} B.,  {Liang} E.-W.,  {Lu} R.-J.,  {Lin} D.-B.,  {Li} J.,   {Li} L.,  2018, \mn@doi [\apj] {10.3847/1538-4357/aabc13}, \href {https://ui.adsabs.harvard.edu/abs/2018ApJ...859..160W} {859, 160}

\bibitem[\protect\citeauthoryear{{Yang} et~al.,}{{Yang} et~al.}{2023}]{2023ApJ...947L..11Y}
{Yang} J.,  et~al., 2023, \mn@doi [\apjl] {10.3847/2041-8213/acc84b}, \href {https://ui.adsabs.harvard.edu/abs/2023ApJ...947L..11Y} {947, L11}

\bibitem[\protect\citeauthoryear{{Zhang} \& {Kobayashi}}{{Zhang} \& {Kobayashi}}{2005}]{2005ApJ...628..315Z}
{Zhang} B.,  {Kobayashi} S.,  2005, \mn@doi [\apj] {10.1086/429787}, \href {https://ui.adsabs.harvard.edu/abs/2005ApJ...628..315Z} {628, 315}

\bibitem[\protect\citeauthoryear{{Zhang} \& {M{\'e}sz{\'a}ros}}{{Zhang} \& {M{\'e}sz{\'a}ros}}{2001}]{2001ApJ...552L..35Z}
{Zhang} B.,  {M{\'e}sz{\'a}ros} P.,  2001, \mn@doi [\apjl] {10.1086/320255}, \href {https://ui.adsabs.harvard.edu/abs/2001ApJ...552L..35Z} {552, L35}

\bibitem[\protect\citeauthoryear{{Zhang}, {Murase}, {Ioka}  \& {Zhang}}{{Zhang} et~al.}{2023}]{2023arXiv231113671Z}
{Zhang} B.~T.,  {Murase} K.,  {Ioka} K.,   {Zhang} B.,  2023, \mn@doi [arXiv e-prints] {10.48550/arXiv.2311.13671}, \href {https://ui.adsabs.harvard.edu/abs/2023arXiv231113671Z} {p. arXiv:2311.13671}

\bibitem[\protect\citeauthoryear{{Zhang}, {Wang}  \& {Zheng}}{{Zhang} et~al.}{2024}]{2024JHEAp..41...42Z}
{Zhang} B.,  {Wang} X.-Y.,   {Zheng} J.-H.,  2024, \mn@doi [Journal of High Energy Astrophysics] {10.1016/j.jheap.2024.01.002}, \href {https://ui.adsabs.harvard.edu/abs/2024JHEAp..41...42Z} {41, 42}

\bibitem[\protect\citeauthoryear{{van Eerten}}{{van Eerten}}{2014}]{2014MNRAS.442.3495V}
{van Eerten} H.,  2014, \mn@doi [\mnras] {10.1093/mnras/stu1025}, \href {https://ui.adsabs.harvard.edu/abs/2014MNRAS.442.3495V} {442, 3495}

\makeatother
\end{thebibliography}




\appendix




\bsp	
\label{lastpage}
\end{document}